\documentclass[twocolumn,showpacs,preprintnumbers,amsmath,amssymb,prl]{revtex4}

\usepackage{graphicx}
\usepackage{dcolumn}

\usepackage[latin1]{inputenc}
\usepackage[T1]{fontenc}
\usepackage{ae,aecompl}

\begin{document}

\newcommand{\rar}{$\rightarrow$}
\newcommand{\lrar}{$\leftrightarrow$}

\newcommand{\beq}{\begin{equation}}
\newcommand{\eeq}{\end{equation}}
\newcommand{\bea}{\begin{eqnarray}}
\newcommand{\eea}{\end{eqnarray}}
\newcommand{\Req}[1]{Eq. (\ref{E#1})}
\newcommand{\req}[1]{(\ref{E#1})}
\newcommand{\degree}{$^{\rm\circ} $}
\newcommand{\pcite}{\protect\cite}
\newcommand{\pref}{\protect\ref}
\newcommand{\Rfg}[1]{Fig. \ref{F#1}}
\newcommand{\rfg}[1]{\ref{F#1}}
\newcommand{\Rtb}[1]{Table \ref{T#1}}
\newcommand{\rtb}[1]{\ref{T#1}}

\title{DNA flexibility on short length scales probed by atomic force
 microscopy}

\author{Alexey K. Mazur}
\affiliation{UPR9080 CNRS, Université Paris Diderot, Sorbonne Paris Cité,\\
Institut de Biologie Physico-Chimique,\\
13, rue Pierre et Marie Curie, Paris,75005, France.}
\author{Mounir Maaloum}
\affiliation{Institut Charles Sadron, CNRS--University of Strasbourg,\\
23 rue du Loess, BP 84087, 67034 Strasbourg cedex 2, France.}

 


\begin{abstract} 
Unusually high bending flexibility has been recently reported for DNA
on short length scales. We use atomic force microscopy (AFM) in
solution to obtain a direct estimate of DNA bending statistics for
scales down to one helical turn. It appears that DNA behaves as a
Gaussian chain and is well described by the worm-like chain model at
length scales beyond 3 helical turns (10.5nm). Below this threshold,
the AFM data exhibit growing noise because of experimental
limitations. This noise may hide small deviations from the Gaussian
behavior, but they can hardly be significant.
\end{abstract}


\maketitle

Long double stranded DNA behaves as a continuous elastic rod with
bending deformations described by the harmonic worm-like chain (WLC)
model \cite{Bresler:39,Landau:76,Cantor:80b}. In many biological
processes the DNA flexibility, notably its ability to wrap around
proteins, plays a key role, therefore, the bendability of DNA is
actively studied \cite{Crick:75,Hagerman:88,Peters:10b}. Recent
experimental data indicate that the WLC model significantly
underestimates the probability of strong bends on length scales
shorter than the persistence length ($l_b$=50nm)
\cite{Cloutier:04,Cloutier:05,Wiggins:06a,Yuan:08,Chen:10e,Vafabakhsh:12}.
This hypothesis is vigorously disputed because it does not agree with
all data
\cite{Du:05a,Mzprl:07,Shroff:08,Linna:08,Demurtas:09,Mastroianni:09,Forties:09,Ortiz:11,Schopflin:12,You:12,Vologodskii:13b}
and because the effect is crucial for biology
\cite{Podgornik:06,Nelson:12,Vologodskii:13a}.

The atomic force microscopy (AFM) has the advantage of directly
observing DNA when adsorbed onto supporting surfaces. Earlier studies
showed that in mild conditions the DNA molecules equilibrate on the
surface by 2D diffusion so that the chain statistics is not perturbed
\cite{Rivetti:96,Faas:09}. Two groups earlier used this method for
studying the statistics of bending in short DNA
\cite{Wiggins:06a,Chen:10e}. It was found that for lengths >30nm the
probability distributions of bend angles and end-to-end distances agree
with the WLC model, but for shorter lengths the populations of strongly
bent conformations are much higher than the WLC predictions.  In
contrast to DNA cyclization, where high probabilities of small circles
can be due to rare fluctuations like melting bubbles
\cite{Altan-Bonnet:03,Yan:04,Zeng:04,Yan:05,Wiggins:05,Ranjith:05,Yuan:06,Destainville:09},
the AFM data suggested that the double helix is intrinsically kinkable,
that is, it is kinked rather than bent smoothly even for small angles.
These results were accounted for by the linear sub-elastic chain (LSEC)
model \cite{Wiggins:06a}.  According to it the bending of DNA fragments
of finite length $l$=2.5nm obeys Boltzmann statistics with an empirical
energy function \beq\label{Elsec} E_{\rm
LSEC}(\theta)=\alpha\left|\theta\right|kT \eeq where $\theta$ is the
bend angle and $kT$ is the thermal energy.  The dimensionless constant
$\alpha$ fit to experimental data equals 6.8. Lengths shorter than 2.5nm
are not considered.

\begin{figure}[ht]
\centerline{\includegraphics[width=8cm]{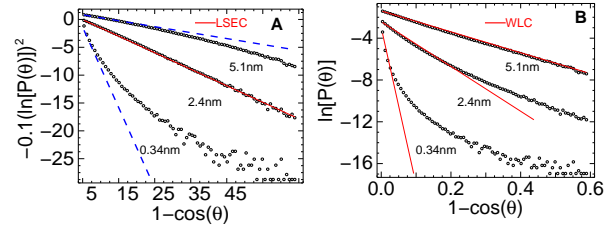}}
\caption{\label{Flsec} Color online.
Probability distributions of DNA bending obtained by BD simulations of
LSEC model. The DNA lengths are shown near the plots. In panel A the
solid red line approximates the LSEC postulate.  The dashed traces are
shown for visual convenience. Solid lines in panel B represent WLC
predictions for $l_b$=50nm.
}\end{figure}

In the earlier literature, one aspect of the LSEC model has escaped
attention. If the neighboring base-pair steps (bps) are approximately
independent the bending probability distribution for $l$=2.5nm
represents a convolution of several single-bps distributions. By
solving the inverse problem  one can derive the effective single-step
potential from \Req{lsec}. A reasonably accurate approximate solution
can be easily found by trials and errors. \Rfg{lsec} displays the
results of Brownian dynamics (BD) simulations with such potential. The
discrete coarse grained model of DNA from the earlier report
\cite{Mzjpc:08} was employed with one bead per base pair.  The
single-bps bending potential was as follows
\beq
U_1(\theta)=\left\lbrace
\begin{array}{cr}
q\theta^2 & \theta<\theta_0\nonumber\\
q\theta_0^2-\frac{q}k\theta_0(\theta_0-\pi)\left[1-
\left(\frac{\theta-\pi}{\theta_0-\pi}\right)^{2k}\right] &
\theta\geq\theta_0
\end{array}
\right.
\eeq
with $\theta_0=2^\circ$, $k=2$, and $q$=110 kcal/mole. It is harmonic
around zero to avoid singularity, but concave beyond a narrow
vicinity, which favors sharp bends. The X and Y-axis scales in
\Rfg{lsec}A and \rfg{lsec}B linearize the bend angle distributions of
LSEC and WLC models, respectively \cite{Mzprl:07}.  The results are
shown for three DNA lengths. For $l=0.34$nm the bend angle distribution
is very broad due to easy kinking. For $l=2.4$nm it agrees with the LSEC
hypothesis (\Rfg{lsec}A), and diverges from the WLC distribution with
large angles (\Rfg{lsec}B). However, for $l=5.1$nm the distribution is
already indistinguishable from the WLC prediction with $l_b$=50nm
(\Rfg{lsec}B). This result demonstrates that \Req{lsec} is not
sufficient to account for the AFM data because the uniform flexibility
of the double helix ensures very rapid convergence to a Gaussian
behavior \cite{Vologodskii:13a}. A much stronger tacit assumption of
LSEC model is that sharp bends are spaced by 2.5nm intervals of straight
DNA \cite{Wiggins:06b}, which is hard to believe.

To shed light upon the above difficulty we used the AFM method to
evaluate DNA bending at short length scales. Linear DNA with fixed
length of 4363 bp was obtained  by cutting PBR322 plasmid with EcoRI
restrictase. Experiments were performed in a solution containing 10 mM
tris-HCl buffer, pH 7.5, supplemented with 1mM MgCl2, to a final DNA
concentration of 1 mg/ml. 200 ml of this DNA solution was injected in
AFM liquid cell and DNA molecules adsorbed onto freshly cleaved
muscovite mica at room temperature. Images were collected using a
Nanoscope 8 (Bruker) operated in tapping mode in solution, with a pixel
size (grid spacing) of 1.95nm.  Ultrasharp non-contact silicon
cantilevers Multi75Al (NanoAndMore) were driven at oscillation
frequencies in the range of 20-26 kHz.  During AFM imaging, the force
was reduced in order to avoid dragging of DNA by the tip. The line scan
rate was usually 1.4 Hz. Integral gain was adjusted to give sharp
images. Images were taken without on-line filtering and were
subsequently processed only by flattening to remove the background
slope. The AFM images of DNA were transformed into discrete chains under
visual control by using a custom implementation of the tracing algorithm
by Wiggins et al \cite{Wiggins:06a}. This procedure was repeated several
times using different link lengths $l_0$. Five independent sets of
contours thus obtained were chosen for further analysis. The
corresponding $l_0$ values were 2.5, 3.5, 7, 10.5 and 14nm,
respectively. The total contour length of DNA observed in the AFM images
was $\sim$348 $\mu$m ($\sim10^6$ bp). With $l_0$=2.5nm this gave about
139,000 angles between adjacent links (compared to 98,000 in the earlier
report \cite{Wiggins:06a}).  For statistical estimates the contours were
divided into fragments so that every measured angle was counted only
once.

The AFM data were compared with Monte-Carlo (MC) simulations of planar
discrete WLC and LSEC models. A phantom chain was considered without
the excluded volume effect. The bend angles were sampled directly from
appropriate Boltzmann distributions.  To get a feel of statistical
errors the DNA length and the volume of sampling were similar to those
in experiment.  The link length $l_0$ in the digitized AFM DNA
contours is always larger than one bps. For short chains this gives a
significant bias with respect to the underlying DNA. To take this into
account, MC simulations of the WLC model were performed with one bead
per bp, but the resulting chain configurations were resampled by
stepping along MC bead positions with fixed strides corresponding to
link lengths in AFM data. These new contours were processed in the
same way as experimental data to generate reference WLC curves.

\begin{figure}[ht]
\centerline{\includegraphics[width=8cm]{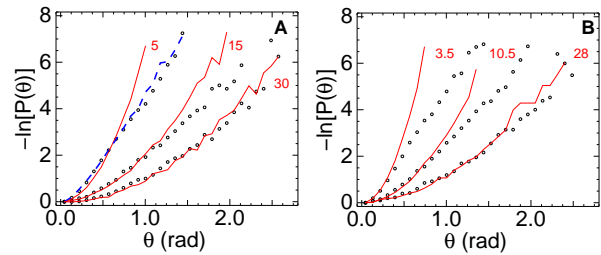}}
\caption{\label{Fafm1} Color online.
Negative logarithm of the probability distribution $P(\theta)$ for the
angle $\theta$ between tangents separated by different contour lengths
indicated in the figure. The AFM images were traced with link lengths
$l_0$=2.5nm (panel A) and 3.5nm (panel B). The dots are experimental
data. Solid lines: MC evaluation of the same function for the WLC
model with $l_b$=56nm. The dashed line in panel A displays the MC
evaluation for LSEC model (\Req{lsec}).
}\end{figure}

Before the beginning of this study we considered two possible origins
of short-length deviations of AFM results from the WLC theory. The
first of them is related to experimental conditions. The short length
deviations from the WLC theory were earlier observed in AFM of dry DNA
\cite{Wiggins:06a,Chen:10e}. During drying, strong DNA-ion
interactions and other electrostatic effects may change  the DNA
conformation. In contrast, solution AFM allows direct visualization
of DNA in nearly physiological conditions \cite{Lyubchenko:97b}. To
check if this experimental difference plays a role we analyzed our AFM
images by a procedure identical to that in the original report by
Wiggins et al \cite{Wiggins:06a}. It turned out that the experimental
estimates of DNA bendability obtained on air and in solution agree
nearly perfectly.  Some representative results are shown in
\Rfg{afm1}A. In agreement with the LSEC model short DNA exhibits
excessive flexibility, but for lengths beyond 30nm everything
converges to the WLC theory. The only notable difference from the
earlier data is a somewhat larger asymptotic $l_b$ value (56nm). This
may be due to solvent conditions, a systematic bias in the measured
DNA length, or the exclusion volume effect \cite{Rivetti:96}.

The second possibility we considered was that the deviations from the
WLC model might be due to the choice of $l_0$=2.5nm in the tracing
algorithm. This length corresponds to 0.7 of a helical turn.  As a
result, the bending is measured only for DNA fragments with
non-integral numbers of turns. It is known from all atom molecular
dynamics simulations that measuring bend angles in such fragments
is prone to large errors due to rotation of reference bp-frames and
anisotropy of bending towards DNA grooves \cite{Mzbj:06,Mzjpc:09}. The
AFM resolution is lower, but this difficulty should persist for any
method that tries to probe bending in DNA fragments of a few helical
turns. \Rfg{afm1}B reveals, however, that qualitatively similar
deviations from the WLC model are evident also when the AFM images are
traced with $l_0$=3.5nm corresponding to one helical turn.

\begin{figure}[ht]
\centerline{\includegraphics[width=8cm]{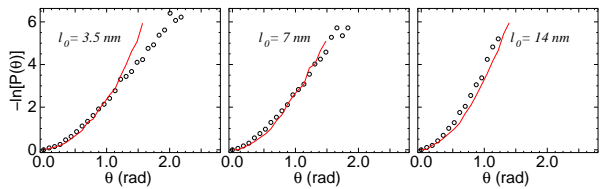}}
\caption{\label{Fafm2} Color online.  Negative logarithm of the
probability distribution $P(\theta)$ for the angle $\theta$ between
tangents separated by a contour length of 14nm.  The AFM images were
traced with three different link lengths indicated in the figure. The
dots are experimental data. Solid lines: MC evaluation of this
function for the WLC model traced with the same link lengths.
}\end{figure}

Continuing the search, we decided to check the consistency of the
results obtained with different link lengths. \Rfg{afm2} compares the
bend angle distributions for 14nm DNA in AFM contours traced with
$l_0$=3.5, 7, and 14nm (1, 2, and 4 DNA turns, respectively). Strong
deviations from the WLC model are observed only with $l_0$=3.5nm. With
$l_0$=7nm they are much smaller and disappear completely with
$l_0$=14nm. For any smooth contour the measured bend angles depend
upon $l_0$ simply due to discretization. As explained above, this
effect is taken into account in the WLC curves in \Rfg{afm2}. With
$l_0$ increased, the shape of the WLC probability distribution is
preserved, but it is uniformly scaled and corresponds to a higher
$l_b$ value (see the three reference WLC curves in \Rfg{afm2} and
discussion below).  In contrast, the experimental data in \Rfg{afm2}
reveal that, with $l_0$ increased, strong bends are suppressed
selectively, and so that the distribution approaches the theoretical
result of the WLC model.

To get further, we systematically checked AFM images that contributed
high populations of strong bends with $l_0$=3.5nm. For instance, in
the molecule in \Rfg{afm3}A there was a kink of 78.6$^\circ$. The
tracing was repeated 30 times in opposite directions starting from
different points. These new contours usually contained a few bends
beyond 43$^\circ$ (the upper limit for the left WLC curve in
\Rfg{afm1}B). However, these bends almost never occurred near the
original kink and the new kink locations varied in repeated contours.
\Rfg{afm3}B shows a fragment of this DNA with a bundle of 30 contours
superimposed. In the AFM images the width of the DNA varied between 4
and 8 pixels \cite{Epaps}. The bundle width in \Rfg{afm3}B also is not
constant and reaches three pixels (5.8nm). Irreproducible strong bends
commonly belonged to zones where the measured DNA width was larger
than average. With $l_0$ increased, the bundle width is reduced.
\Rfg{afm3}C shows the results of similar tests with $l_0$=14nm.

\begin{figure}[ht]
\centerline{\includegraphics[width=8cm]{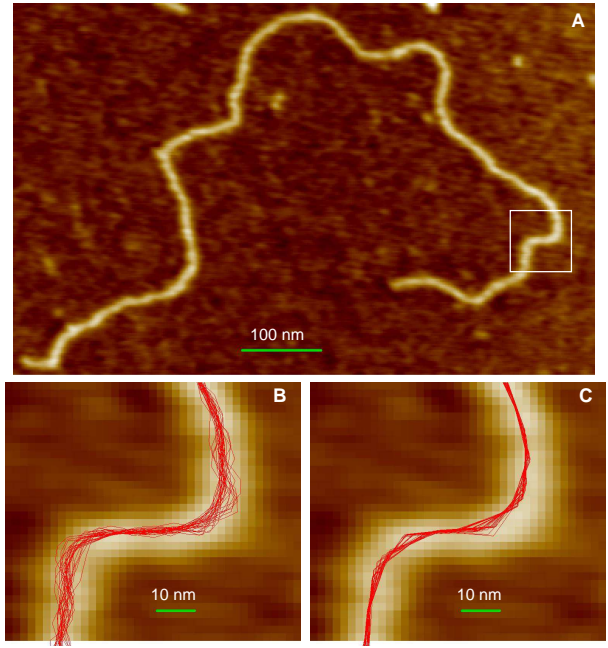}}
\caption{\label{Fafm3} Color online.
Panel A: a high resolution AFM image of a DNA molecule selected as
described in the text. It was traced 30 times using two alternative
directions and starting from arbitrary points at opposite ends. Panel B
shows a zoomed out view of the area highlighted by the white rectangle
including a fragment of DNA with the bundle of contours computed with
the link length $l_0$=3.5nm. Panel C shows similar results for tracing
with a larger link length $l_0$=14nm.
}\end{figure}

The above results demonstrate that the measured population of large
bending angles rapidly grows when the $l_0$ value is reduced to
lengths comparable to the DNA width in AFM images. \Rfg{afm2} and
\rfg{afm3} strongly suggest that this effect represents a limitation
of data processing rather than the physical property of DNA. A
sensible comparison with the WLC theory is possible only when this
effect is small. Based upon \Rfg{afm2} and \rfg{afm3} we concluded
that the appropriate link length for our data is $l_0$=7nm (two
helical turns). With larger link lengths the contours are increasingly
prone to round sharp bends (see \Rfg{afm3}C). Some deviations from the
WLC model are still evident with $l_0$=7nm, but they are not very
significant. The link length $l_0$=10.5nm was also checked, and this
gave results similar to $l_0$=14nm, that is, deviations from the WLC
model were absent, but the apparent $l_b$ values were slightly
overestimated due to cutting sharp bends. \Rfg{afm4} shows a more
detailed analysis of the shape of the probability distribution
functions obtained with $l_0$=7nm in the range of small angles where
the data can be linearized by an appropriate choice of scales.

According to the WLC model, DNA of contour length $L$ equilibrated on
a plane is described by the normalized bend angle distribution
\cite{Rivetti:96}
\beq\label{Ewlc1}
P\left[\theta(L)\right]_{2D}=\sqrt{\frac{l_b}{2\pi
L}}\exp\left(-\frac{l_b \theta^2}{2L}\right)
\eeq
so that
\beq\label{Ewlc2}
\langle\theta^2\rangle=L/l_b.
\eeq
For small angles \Req{wlc1} gives linear plots in coordinates
$[\sin(\theta)P(\theta)]$ versus $[1-\cos(\theta)]$, for instance.
\Rfg{afm4}A shows the results of MC simulations of a WLC model with one
bead per bp traced with the link length $l_0$=7nm. The distribution
functions for the three smallest contour lengths all have linear shapes
corresponding to the WLC model. The apparent $l_b$ value is visibly
overestimated due to discretization. As seen in \Rfg{afm4}C, with $L$
increased, $l_b$ decreases, but does not reach the value of the
underlying WLC model (56nm) \cite{Epaps}. \Rfg{afm4}B and \rfg{afm4}D
display similar plots for experimental AFM data. As expected, for
$L$=7nm the shape of the probability distribution slightly deviates from
the WLC model, nevertheless, the overall pattern is evidently similar to
that in MC.

\begin{figure}[ht]
\centerline{\includegraphics[width=8cm]{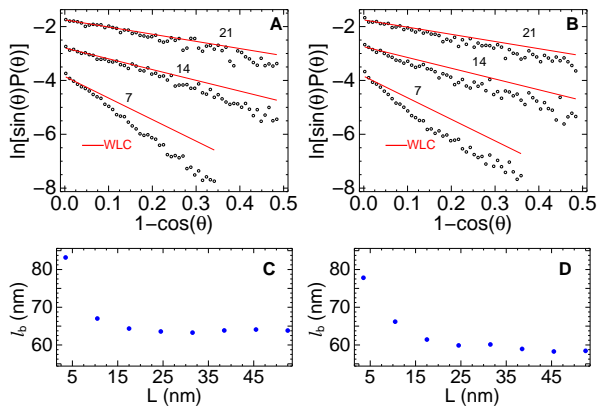}}
\caption{\label{Fafm4} Color online.
Probability distributions of DNA bending obtained with $l_0$=7nm for MC
simulations (panels A,C) and experimental data of AFM (panels B,D). The
plots are labeled by the corresponding DNA lengths (in nm). Straight
lines in both panels represent WLC predictions for $l_b$=56nm. Panels C
and D display the values of the bending persistence length $l_b$
obtained with \Req{wlc2}
}\end{figure}

The non-Gaussian statistics of bending fluctuations in short DNA were
observed in two earlier AFM studies \cite{Wiggins:06a,Chen:10e}, and
also here in \Rfg{afm1}. The experimental conditions and parameters of
DNA images in these three cases were not identical, but the linear
contours were obtained with the same tracing algorithm. Transforming
AFM data into linear contours is not trivial because there is no
constructive definition of the centerline of a DNA image.  Defining it
as a minimum-cost path, for instance, leads to biased contours with
underestimated flexibility \cite{vanHeekeren:07}. The tracing algorithm
by Wiggins et al \cite{Wiggins:06a} is rapid and simple. It uses a
manually set starting point and search direction, therefore, its
result is a bundle of contours rather than a single line. We found
that the spread of this bundle dramatically grows when the tracing
link length is reduced below a certain limit. This effect was
responsible for the apparent deviations from the Gaussian behavior in
our experiment.  When it is negligible the measured bending statistics
agrees with the WLC model, which involves DNA lengths beyond three
helical turns. The origin of non-Gaussian effects in the earlier
studies \cite{Wiggins:06a,Chen:10e} could be different because they
employed AFM in air and parameters of DNA images were different. We
believe that AFM in solution can be also used for probing smaller DNA
lengths, but this should require additional work on experimental
conditions and data processing.


In summary, AFM experiments demonstrate that bending fluctuations in
DNA absorbed on a plane in solution are Gaussian and well described by
the WLC model at all length scales beyond 3 helical turns (10.5nm). With
DNA lengths reduced below this threshold, the AFM data exhibit growing
noise because of experimental limitations. This noise may hide small
deviations from the Gaussian behavior, but they can hardly be significant.

\begin{acknowledgments}
\end{acknowledgments}

\bibliography{last}

\end{document}